# Draft of technical paper

# Safety design concepts for statistical machine learning components toward accordance with functional safety standards

Akihisa Morikawa[1]  and Yutaka Matsubara[2]

[1] Witz Corporation, Aichi, Japan
morikawa@witz-inc.co.jp
[2] Nagoya University, Aichi, Japan
yutaka@ertl.jp

**Abstract.** In recent years, curial incidents and accidents have been reported due to unintended control caused by misjudgment of statistical machine learning (SML), which include deep learning. The international functional safety standards for Electric/Electronic/Programmable (E/E/P) systems have been widely spread to improve safety. However, most of them do not recommended to use SML in safety critical systems so far. In practical the new concepts and methods are urgently required to enable SML to be safely used in safety critical systems.
  In this paper, we organize five kinds of technical safety concepts (TSCs) for SML components toward accordance with functional safety standards. We discuss not only quantitative evaluation criteria, but also development process based on XAI (eXplainable Artificial Intelligence) and Automotive SPICE to improve explainability and reliability in development phase. Finally, we briefly compare the TSCs in cost and difficulty, and expect to encourage further discussion in many communities and domain.

## 1      Introduction

In recent years, AI technology has been widely used to improve performance and/or efficiency of functionality in safety-critical systems, and realizes automation with many input data in complex situations. The safety critical services including automated driving are drastically changing as well. For autonomous driving systems, AI technology especially SML is employed in safety critical functions. However, many cases that lead to crucial accidents due to unintended control caused by SML misjudgment have been reported [1]. In general, SML systems are implemented as an E/E/P system. The international functional safety standards for E/E/P systems have been widely used in various domains. IEC 61508 [2] for general, ISO 26262 [3] for automobiles and ISO 13849 [4] for control systems are typical. Unfortunately, safety of E/E/P systems using SML are not completely argued in the standards at this time, and then dangerous incidents and accidents regarding the unintended behaviors of SML systems occurred in practical situations. Therefore, new safety concepts and methods for SML systems are urgently required in many industries.

In TIGARS project [5], we have widely studied gaps between development of conventional deductive systems and one of SML systems especially for autonomous vehicles. In SEAMS project [6], we have researched safety argumentation and measures for safety-related SML systems. We have also developed typical design concepts and learning processes of SML to comply with functional safety standards, and finally published the guideline document as *The safety design guidelines for SML systems,* which is currently in Japanese only. Based on our guideline, in this paper, we discuss TSCs for safety-related SML systems toward accordance with functional safety standards, and compare them in cost and difficulty for practical developments. We also assume that HARA (Hazard Analysis and Risk Assessment), specification of safety goals, which are top-level safety requirements (SRs), and assignment of SRs to the target SML system have already been finished, then discuss how the SRs can be achieved in concept, design, implementation and verification phases, which correspond to typical V-model shown in IEC 61508 and ISO 26262. The one of the popularly-implemented safety measures for functional safety is an additional safety monitoring function for safety related components. Then, the SML component can be treated under QM (Quality Management). In addition to the mechanism, we also propose TSCs, where no additional mechanism is required, and consider quantitative criteria applicable for SML components with SIL (Safety Integrity Level)/ASIL (Automotive SIL).

The contributions of this paper are as follows.
- We analyze gaps and make technical issues clearer for SML components toward accordance with functional safety standards.
- We present TSCs to mitigate or solve the technical issues.
- We propose safety measures to ensure both reliable behaviors and reliable development process for safety-related SML components.
- We briefly compare the proposed TSCs in cost and difficulty for practicality.

## 2    Related research

Current functional safety standards do not recommend the use of AI including SML, but several documents have been published to consider safety of AI. Table A.2 of IEC 61508-3:2010 and Annex G of SOTIF standard [7] present an off-line training process for SML, but these documents do not mention specific criteria or safety design policies. European Commission [8][9] describes management measures for SML. However, these documents do not mention any concrete safety design for functional safety. Many studies have examined safety designs for safety-related SML systems such as autonomous driving. Mobileye's approach [10] is to set rules that prevent the automated driving system from producing accidental output. DeepMind [11][12] proposed a method for repeating reinforcement learning to ensure that SML systems output only in safety range. However, neither mentions functional safety standards. As described above, there are currently no technical documents that specify concrete requirements for safety-related SML systems to comply with functional safety standards.

Mentor's White Paper [13] shows architectural design for safety-related SML systems and expects that the safety criteria will be introduced in the newest version of [7], which is considered as preprocess of functional safety activities. However, it does not mention concrete design and evaluation of SML software component. Rick [14] shows technical problems for AI to comply with ISO 26262, but did not mention specific measures to solve them. Leading companies for autonomous driving in Germany, the United States, and China published *Safety First for Automated Driving* [15] for dependability including both of safety and security. This document is under working in ISO TC 22 to be published as TR4804. In Chap. 6 of the document, issues, expected work products, and activities in learning are considered for the development process of safety-critical components using supervised DNN (Deep Neural Network). However, no clear correspondence with the functional safety standards and no quantitative evaluation criteria is described.

In ELISA project [16], functional safety-compliant measures for Linux kernel, which is one of the popular and complex software platforms, have been discussed. The quality of such complex and deductive software can be theoretically assured with the white box approach required in functional safety standards. However, it is difficult to take same approach for SML systems because of its inductive behavior. The technical issues are explained in the next section.

In recent years, highly explanatory software development based on Automotive SPICE [17] has become popular in the automotive industry for sufficient reduction of risks on development and accountability for development process quality. The new standards such as ISO 25119 [18] for agricultural machinery and ISO 19014 [19] for construction machinery have been also published. However, it is difficult to show that the risk is sufficiently reduced since the learning process of SML is not included in such standards. Patrick [20][21] proposes workflows and guidelines for AI development. However, it did not clarify the details of each process and not sufficiently explained about risk reduction measures required in functional safety standards.

## 3  Proposed technical safety concepts

### 3.1  Technical issues

SML is typically established as neural network form, and predicts inductively after recursive learning of its model. AI including SML is not now recommended to be used for fault correction due to its dynamic reconfiguration according to Table A.2 of IEC 61508-3:2010. For more details, we believe that followings especially regarding SML systems are not allowed or strongly prohibited in the current functional safety standards.
-     SRs including probabilistic behaviors

- Design and implementation with explicit probability of fails to meet SRs
- Recursive learning as dynamic reconfiguration

Since deductive systems work as designed, the behaviors of them can be verified and be reliable through reliable development process based on practical standards or guideline. On the other hand, inductive systems like SML systems are developed through recursive learning using finite training dataset. Supervised SML outputs prediction results, which may be probabilistically calculated on its neural network. Unsupervised SML outputs a probabilistic distribution or a decision result based on that. Therefore, outputs from SML can be wrong even if its design and implementation are completely performed through reliable process. According to this essential reason, we believe that development of SML cannot directly meet requirements in functional standards, and then new safety concepts for SML systems are necessary.

We assume that prediction algorithms for SML is implemented in software through functional safety development process. A typical software development process like V-model in functional safety standards consists of two kinds of activities. One is to ensure reliability of software behaviors (outputs), and another one is to ensure quality of reliable process, where design, implementation, verification and validation based on SRs assigned to the component are included. We analyzed gaps between development process for conventional deductive software components and one for SML software components. In Table 1, technical issues for key activities in software development process are described. We believe that most of them are similarly applicable to hardware component development.

**Table 1.** Key activities and technical issues for SML software development

| Key activities | Characteristics of SML | Technical issues |
| --- | --- | --- |
| [P-a] SR specification | The prediction result (output) differs possibly depending on input and external environment. If the prediction employs stochastic logics, the determinism of the behavior becomes quite low. | It is difficult to clearly specify requirements including preconditions, inputs and outputs. The test specifications extracted from SRs cannot comprehensively specified. Then, it is difficult to ensure sufficient quality of the output. |
| [P-b] Consistency checking between requirements and implementations | In general, to understand the internal structure and parameters of SML is quite difficult for engineers. Then, keeping consistency among requirements, design and implementation of SML is quite difficult as well. | It is quite hard to investigate the root cause of the incorrect output. The impact analysis of retraining of the model cannot be done with fine grain. This means that whole of the SML system must be revalidated after any retraining. |
| [P-c] White-box verification | Same as the above. | The exhaustive white-box test and review are difficult in unit and integration verification phase. If such phases are skipped, it is suspicious whether sufficient quality is ensured only by system integration test and validation. |

| | | | |
|---|---|---|---|
| [P-d] Quality management process | The learning methods for SML are currently diverse and flexible. | | No standardized process to mitigate or avoid high risk during learning and re-learning phases. |

## 3.2 Technical safety concepts for safety-related SML systems

To solve technical issues for SML, we propose the five types of TSCs shown in Table 2. For each concept, we describe relevant documents and qualities to be assured for the target SML system or software component. More details are described in the following sections.

Table 2. TSCs for safety related SML software components

| TSCs | | Relevant documents | Safety assurance in system level | Safety assurance in software component level | |
|---|---|---|---|---|---|
| | | | | Reliable process | Reliable output |
| (1) Safety design for safety-related SML Systems | (1-1) Safety monitoring mechanism | Sec.7.4 of IEC 61508-2:2010, Sec.6 of ISO 26262-4:2018, Sec.6.2.5 Cat.2 of ISO 13849-1:2015 | X | N/A | N/A |
| | (1-2) Redundant SML software components | Figure B.2 of ISO 26262-5:2018 | N/A | N/A | X |
| (2) functional safety capable development process for safety-related SML software component | (2-1) Assessment of SML development process | Sec.7.4.2.12 and Sec.7.4.2.13 of IEC 61508-3:2010, Sec.6.4.12 of ISO 26262-2:2018 | N/A | X | N/A |
| | (2-2) Quantitative evaluation of dangerous failure rate | Sec.5 of IEC TS 62998-1:2019 | N/A | N/A | X |
| | (2-3) Proven in use | Sec.7.4.2.12 of IEC 61508-3:2010, Sec.14 of ISO 26262-8:2018, Table 7 of ISO 13849-1:2015 | N/A | X | X |

## 3.3 (1-1) Safety monitoring mechanism

**Explanation:** As an additional safety mechanism, a monitoring system is introduced to mitigate or avoid dangerous situation outside the SML software component. When any malfunctions happen at the component, the safety mechanism detects dangerous situation, and then stops the software or output a safe value as priori to ensure safety of whole of the system. This is one of the most popular safety measures for functional safety. To detect abnormal and dangerous situation, just as examples, output of SML component or feedback from actuator should be checked.

**Requirements for functional safety standards:** The sufficient independence must be established between the SML component and the safety monitoring mechanism according to Sec.7.4.2.3 of IEC 61508-2:2010. Both of common cause failures and cascading failures between them are comprehensively analyzed by system FMEA (Failure Mode and Effect Analysis), and fundamental causes are removed. Then, the SML component can be treat as non-safety related one, and the safety mechanism should be developed with SIL/ASIL assigned to the SML component originally.

### 3.4  (1-2) Redundant SML software components

**Explanation:** Two or more redundant components of the SML software component are additionally introduced to reduce probability of fails to meet SRs. This concept is based on safe fault, which is defined in ISO 26262-1:2018 and explained in Figure B.2 of ISO 26262-5:2018. In ISO 26262, safe fault is explained as one of failure mode classifications of a hardware element. The three or more multiple faults that occurs independently are considered as a safe fault, and no safety measures are required for them. No quantitative criteria like failure rate of parts is mentioned. We believe that it is possible that unintended output of three or more redundant SML software components can be considered as a safe fault if these components work independently and probability of fails to meet SRs is not highly increased.

**Requirements for functional safety standards:** For liable output, three or more redundant components of the SML component are differently designed and allocated to sufficiently independent elements. Each component should work as self-contained one with safety mechanisms, which detect faults, errors, or failures regarding itself, and transit to safe state and stop its output to avoid any interference to other components. We believe that function-level or design-level diversity is required. For functional-level diversity, each component uses different kinds of inputs and have intended functionalities for the component. For design-level diversity called N-version design, each component employs different SML model, learning data and be trained by different development team.

For liable design, functional-level or design-level diversity contributes to obviously decrease systematic failures as well. However, as the best of our knowledge, no concepts or methods is introduced to reduce SIL/ASIL assigned to the SML component after HARA phase except for ASIL decomposition in ISO 26262-9:2018. Therefore, safety measures for systematic failures in each redundant component may be required based on SIL/ASIL originally assigned to the SML component. (2-1) can be additionally considered as safety measures. While unintended output from SML is considered as a result due to performance limitations in [7], in this concept, it is considered like component failure.

47

## 3.5 (2-1) Assessment of SML development process

The technical issues of SML are listed in Table 1. To solve these issues towards accordance with functional safety standards, we discuss whether the XAI techniques can be utilized in Sec.3.5.1. The XAI named by the Defense Advanced Research Projects Agency (DARPA) in the United States is one of modern research topics, and aims to explain clearly prediction results and internal of AI models. We propose an explainable learning process of SML based on Automotive SPICE in Sec.3.5.2.

**Table 3.** Capability of XAI techniques for key activities

| XAI techniques | [P-a] | [P-b] | [P-c] | [P-d] |
|---|---|---|---|---|
| LIME | N | N | P | N |
| CORELS | P | P | P | N |
| Prototype Selection | P | P | N | N |
| MMD-critic | P | P | N | N |
| Explainable SML learning process (Sec.3.5.2) | N | N | N | F |

### 3.5.1 Realization of functional safety process using XAI technology

**Explanation:** In recent years, the XAI techniques have been proposed and introduced in [22]. In Table 3, we focus on several methods to explain internal of AI models or test them, and briefly evaluate them in capability for each key activity, where F, P and N means fully capable, partially capable and non-capable, respectively. LIME [23] explains concretely how outputs from AI components are generated under specific conditions. We believe that white-box tests of SML can be partially performed using LIME, however, exhaustive test is difficult because of incredible number of conditions and situations in general. CORELS [24] searches solutions quickly in combinatorial optimization problems by learning the rule list based on the decision tree, where requirement specifications of SML are clearly described. We believe that [P-a] can be performed by utilizing the rule list if requirements are described as similar optimization problems. While it is difficult to obtain exhaustive traceability among requirements, designs and implementations in SML development, highly reliable tools may be able to support keeping consistency between the rule list and the model of SML component. Unfortunately, such tools do not exist at this time. For classification problem with supervised learning, Prototype Selection [25] finds training data that represents each category. MMD-critic [26] is also helpful to output of SML by presenting both typical and atypical examples for each category. However, it is difficult to validate outputs and weight values in SML.

**Requirements for functional safety standards:** As shown in Table 3, no complete method, which can be capable for all key activities, is proposed. However, combining several XAI techniques may be able to perform partially these activities. The one of critical problems is lack of exhaustive testability because the traditional test methods based on equivalence classes and boundary value analysis cannot be applied to SML. This means that [P-a] is a fundamental issue in SML. Just one possibility, we think that the rule list from the CORELS can be used as preliminary requirement specifications and then equivalence classes and boundary value analysis are also performed. If LIME is also combined, [P-a], [P-b] and [P-c] may be able to be performed better. The development of reliable support tools for these issues are necessary. More detail study regarding this topic is a future work. For [P-d], we believe that explainable SML learning process is required while XAI can be partially helpful for [P-a], [P-b] and [P-c].

### 3.5.2 Explainable SML learning process

**Explanation:** We propose an explainable SML learning process based on Automotive SPICE, which is the de-fact standard in automotive software engineering, to mitigate risks introduced in SML learning process. We tailored the major practices of Automotive SPICE and considered adequate activities for each practice. In Table 4, the part of our proposal is shown.

**Requirements for functional safety standards:** We recommend the developers and engineers of the SML component to record all evidence for each practice, and to explain that all practices have been properly performed based on the proposed process to independent authorities.

Table 4. Example of tailoring Automotive SPICE to the SML learning process (Excerpt)

| Automotive SPICE Version 3.1 Base Practices | Tailoring results of SML learning process |
|---|---|
| SWE.2.BP1: Develop software architectural design. | The design documents for SML architecture are created to clearly define the technical policies like number of layers, type of optimizer type of learning data etc. |
| SWE.2.BP2: Allocate software requirements. | If the SML component is split into multiple elements, the requirements for the component are explicitly assigned to the elements. |
| SWE.2.BP3: Define interfaces of software elements. | If the SML component is split into multiple elements, the interfaces among them are clearly defined. |
| SWE.2.BP4: Describe dynamic behavior. | The dynamic behavior and resources consumption of the SML component are clearly defined and explained. |
| SWE.2.BP5: Define resource consumption objectives. | |
| SWE.2.BP6: Evaluate alternative software architectures. | The design policy is evaluated. |
| SWE.2.BP7: Establish bidirectional traceability. | For the SML component, traceability between requirements specification and architecture design is documented to ensure their consistency. |
| SWE.2.BP8: Ensure consistency. | |
| SWE.2.BP9: Communicate agreed software architectural design. | The architecture design for the SML component is discussed and agreed by developers and engineers when the design is changed. |

### 3.6 (2-2) Quantitative evaluation of dangerous failure rate

**Explanation:** We propose a quantitative evaluation method for SML software component based on IEC TS 62998-1:2019 [27], which is for safety-critical sensor systems and describes a quantitative criteria of dangerous failure rates for each SIL. In [27], the probability of dangerous output from the sensor systems is calculated based on the probabilistic distribution of the sensor's output, and reduction of probability for dangerous output is required according to SIL assigned to the sensor system. Similar to variation of sensor's output under diverse environment, the SML component also depends on environment and non-deterministically behaves. Therefore, we believe that the quantitative criteria can be utilized for SML component. An example of applying the evaluation method in [27] to a distance measurement using SML is shown in Fig. 1.

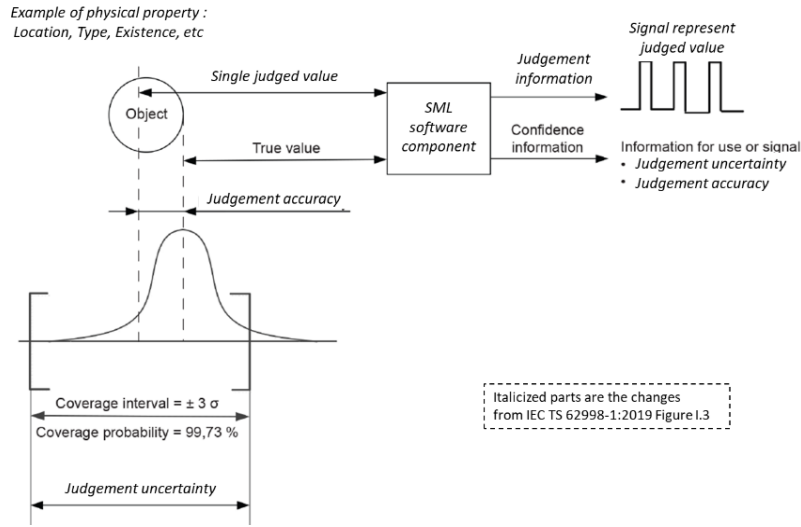

**Fig. 1.** Application of IEC TS 62998-1 to SML

**Requirements for functional safety standards:** The evaluation formula for acceptable probability of the SML component is presented as (1). The right of (1) depicts measurement probability, where L is the upper limit PFH (average frequency of dangerous failure per hour), and D is operation request rate.

$$P > 1 - \frac{L}{D} \quad (1)$$

In case that distribution of outputs, which are distance values to the object in this example, is on the normal distribution, P in (1) is the safety allowable range. Based on the SIL assigned to the component, the target performance to be achieved becomes clear since the performance class is defined for each SIL in Table 1 of [27]. For example, SIL 3 corresponds to the performance class E, and therefore, the minimum required performance is $1-2.5\times10^{-8}$ obtained from Table 5. The safe output range is set as σ of the normal distribution, where σ means the acceptable error range of the SML outputs. When σ is set to a wider value and then P also increases, it is more difficult to achieve the required performance to the SML component. This method is greatly helpful to evaluate performance of SML components.

Table 5. The quantitative evaluation criteria from Table 5 of [27]

| Performance class | A | B | C | D | E | F |
| --- | --- | --- | --- | --- | --- | --- |
| The right side of (1) | $>1-2.4\times10^{-3}$ | $>1-1.0\times10^{-5}$ | $>1-3.0\times10^{-6}$ | $>1-2.5\times10^{-7}$ | $>1-2.5\times10^{-8}$ | $>1-2.5\times10^{-9}$ |

### 3.7 (2-3) Proven in use

**Explanation:** In fact, the concept of proven in use has been accepted in many safety standards. Due to the story, we believe that prove in use is acceptable to utilize SML components in safety critical domains.

**Requirements for functional safety standards:** The acceptable conditions in proven in use is strictly defined. The binary of software and chip product of microprocessor should be completely same, and the history record of a long-term usage, for example one year or longer in IEC 61508, is required as well. In C.2.10.1 of IEC 61508-7:2010 and Sec.14 of ISO 26262-8:2018, the quantitative criteria are defined, such as period of use, failure rate, number of operations, etc. In practical, acceptable conditions explained above can prevent SML components from rapidly improving by retraining of the model and introducing new hardware chips or techniques for SML.

## 4 Discussion

In Table 6, we briefly compared the TSCs explained in this paper, and evaluated them in both of cost and difficulty. The value of 0 (no or nothing), 1 (few or easy), 2 (conventional or normal) or 3 (significant or difficult) is roughly assigned as relative evaluation to each result. In the Total row, total evaluation score for each concept is presented, where lower is better. (1-1) is most applicable in practical development, however it requires additional monitoring systems, which can increase hardware costs. In (1-2), since the failure rate of each hardware element in safety critical systems is assumed to be quite low the idea of safe fault can be acceptable. In case of SML components, it is a technical problem whether failure (i.e. false positive/negative prediction) rate can be decrease similar to hardware elements in practical. As described in Sec.3.5, (2-1) is currently difficult to fully comply with functional safety standards. If that is achievable in practice, future benefits increase. In (2-2), we introduced a well-established evaluation formula. In practice, balancing between performance requirements and safety allowable range should be considered more detailed. As another issue, because [27] is not a popular technical specification at this time, its applicability for SML systems should also be discussed in each industry. (2-3) is the cheapest and easiest, however requires long-term usage and have strict conditions, which can cause a crucial disadvantage in practical system as explained in Sec.0. Through the discussion here, we believe that (2-1) and (2-2) are attractive and practical concepts to reduce development costs in many industries.

Table 6. Brief comparisons for TSCs

| TSCs | Phase | Workloads in newly development | Hardware cost | Realization difficulty | Workloads after re-learning | Total |
|---|---|---|---|---|---|---|
| (1-1) | system architecture design | almost same as conventional deductive systems (2) | additional monitoring system required (3) | same as conventional deductive systems (1) | reintegration test (1) | 7 |
| (1-2) | system architecture design | three or more redundancy with sufficient independence required (3) | three or more redundancy with sufficient independence required (3) | diverse design (3) | component test and reintegration test (2) | 11 |
| (2-1) | development and change management of SML component | almost same as conventional deductive systems (2) | no additional hardware required (0) | more improvements in XAI techniques expected (3) | reassessment of the relearning process (2) | <u>7</u> |
| (2-2) | development and change management of SML component | output measurement and quantitative evaluation required (1) | no additional hardware required (0) | lack of experience (2) | remeasurement of output and reassessment (2) | 5 |
| (2-3) | one or more year(s) after product release | data recording and its evaluation (1) | no additional hardware required (0) | easy but usage condition is restricted (1) | relearning is not allowed (3) | 5 |

## 5  Conclusion

We presented five kinds of TSCs for SML components toward accordance with functional safety standards. We also discussed not only quantitative evaluation criteria, but also development process to improve their explainability and reliability in development phase. Through the discussion above, we concluded that combination of XAI techniques and quantitative evaluation method based on [27] can partially solve technical issues and are practical concepts to reduce development costs at this time. However, further researches and discussions are necessary to completely solve the issues mentioned in this paper. We will continuously study the topics and update our guideline to encourage further discussion in many communities and industries.